# The Striking Impact of Natural Hazard Risk on Global Green Hydrogen Cost

Maximilian Stargardt[a, b, *,] Justus Hugenberg[a], Christoph Winkler[a, b], Heidi Heinrichs[a], Jochen Linßen[a], Detlef Stolten[b]

[a)] Forschungszentrum Jülich GmbH; Institute of Climate and Energy Systems – Jülich Systems Analysis (ICE-2) 52425 Jülich, Germany

[b)] RWTH Aachen University, Chair for Fuel Cells. Faculty of Mechanical Engineering, 52072 Aachen, Germany

* Corresponding author: Maximilian Stargardt

## Abstract

Due to climate change, natural hazards that affect energy infrastructure will become more frequent in the future. However, to incorporate natural hazard risk into infrastructure investment decisions, we develop an approach to translate this risk into discount rates. Thus, our newly developed discount rate approach incorporates both economic risk and natural hazard risk. To illustrate the impact of including the risk of natural hazards, we apply country-specific discount rates for hydrogen production costs. The country-specific relative difference in hydrogen generation cost ranges from a 96% surplus in the Philippines to a -63% cost reduction in Kyrgyzstan compared to a discount rate that only consists of economic risks. The inclusion of natural hazard risk changes the cost ranking of technologies as outcome of energy system models and thus policy recommendations. The derived discount rates for 254 countries worldwide are published in this publication for further use.

## Highlights

- Method to translate country-specific natural hazard risk in investment risks
- Application of developed discount rate approach to country-specific hydrogen cost
- Hydrogen cost change from -63% to +96 % compared to economic risk approach
- Saudi Arabia a stable country to natural hazard risk incorporation
- Provision of 254 country-specific discount rates globally

**Abbreviations:** LCOE – Levelized Cost of Electricity, LCOH – Levelized Cost of Hydrogen, WACC – Weighted average cost of capital

# 1 Introduction

Renewable-based hydrogen can be used as an energy carrier and especially as seasonal storage medium within future decarbonized energy systems [1]. Since hydrogen or its derivatives can act as a long-term storage medium for renewable electricity, it is neither necessary to generate renewable electricity close to demand locations, nor at the same time of demand. Instead, there may be longer distances between supply and demand regions for hydrogen, forcing research to identify preferable regions globally that can produce hydrogen from renewable energy at low cost. Hence, studying levelized cost of hydrogen (LCOH) across the world has been the subject of numerous analyses in the scientific literature. Many studies focus on natural hydrogen supply, for example in Niger [2], Turkey [3], Australia [4], Oman [5], Mexico [6], Uruguay [7] and Egypt [8], regional hydrogen supply [9], [10], or calculate the LCOH for hydrogen that is determined for an export from selected countries [11]. While a substantial amount of scientific work has been done addressing global renewable energy potentials and, therefore, determining full load hours for potentially installed power generation technologies [12], the consideration of region-specific investment risk in levelized cost of electricity (LCOE) and LCOH is underrepresented in research. A well-established method to incorporate investment risks is done through adjusting the discount rate for annualizing total costs with high sensitivities when changing the discount rate [13]. For example, Egerer et al. [14] proved for a global ammonia supply chain from Australia to Germany, that the discount rate is the most sensitive parameter for the LCOH. An overview of the applied discount rates in different research is shown in Table 1.

| Author | Uniform | Country-specific | Technology-specific | Damodaran approach | Source |
|---|---|---|---|---|---|
| Agutu et al. | | x | | | [15] |
| Bachner et al | | x | | | [16] |
| Egli et al. | | x | | | [17] |
| Kan et al. | | x | | x | [18] |
| Kenny et al. | x | | | | [19] |
| Kigle et al. | | x | | x | [20] |
| Mentis et al. | x | | | | [21] |
| Pieton et al. | | x | | x | [22] |
| Polzin et al. | | x | x | | [23] |
| Szabó et al. | x | | | | [24] |
| Wolf et al. | | x | | x | [25] |

*Table 1: Overview of applied global, technology- and country-specific discount rates in literature*

Previous studies have often assumed a uniform discount rate for global or regional calculations, neglecting any differences in investment risks for renewable energy supply chains. For instance, Mentis et al. [21] develop an electrification tool, covering the whole Sub-Saharan Africa, by applying a uniform discount rate of 8%. Kenny et al. [19] compare the ammonia production, and thus the hydrogen generation as an input in the supply chain, from renewable energy in Germany with import scenarios from Chile, Morocco and Namibia. A uniform discount rate of 8% is applied as well. Despite that, Szabó et al. [24] assume a discount rate of 5% to identify economically viable regions for generating electricity from PV in Sub-Saharan Africa, East Asia, and South Asia.

In contrast, Agutu et al. [15] compare the uniform discount rate with country-specific discount rates from public sectors and mainstream financing conditions and show the important implications for electricity grid expansions in Sub-Saharan Africa. According to their results, the cost of capital for off-grid electrification is 32.2% higher with country-specific input parameters, including the discount rate. Bachner et al. [16] emphasize the need to incorporate different investment risks into the discount rate, as this can reduce costs by up to -30% when comparing a uniform discount rate with their base scenario of differentiated discount rates. According to this perspective, it is crucial to adopt regionally differentiated discount rates to provide reliable decision support. Besides a country-specific differentiation, Bachner et al. also adjust the discount rate to different technologies.

The cost of capital and, thus, the included discount rate, is analyzed by Egli et al. [17]. They compare LCOE for different technologies globally and find that the levelized cost of electricity of photovoltaics is 7 - 30% lower for their analyzed developed countries and up to 130% higher for their analyzed developing countries when compared to an application of a uniform discount rate of 7%. Polzin et al. [23] provide a comprehensive research on equity and debt for most European countries and, thus, calculate the WACC for different electricity generation technologies for these countries resulting in WACC ranging from 1.4% to 23.9% across renewable and fossil-fuel based electricity generation technologies. The WACC could be used as discount rate.

To determine country-specific risk rates and, thus, the discount rate, literature often refers to the Damodaran approach [26]. Thus, Kan et al. [18] examine the impact of Damodaran's values from 2021 on the LCOE of electricity generation from photovoltaics in Africa compared to a uniform discount rate, emphasizing the need to differentiate the discount rate. Damodaran's values from 2021 are also the basis of Kigle et al.'s [20] results. They provide global LCOH in a 50 km x 50 km grid utilizing PV and onshore wind power plant in hybrid systems for electricity generation and, further on, for hydrogen generation. Wolf et al. [25] process Damodaran's approach in an analysis that determines the potential LCOH in Germany, Spain, Algeria, Morocco, Norway and Egypt for 2050 by conducting a Monte Carlo simulation. They find that electricity from photo voltaic is the better choice for hydrogen production in all mentioned countries besides Norway.

Applying differentiated discount rate does not only affect LCOE and LCOH but can change the outcome of modeled energy transformation paths, too. For example, Ameli et al. [27] note that lower discount rates are beneficial for reaching climate neutrality earlier than expected. In addition, Emerling et al. [28] combine the application of different discount rates and the carbon emission pricing in their numerical integrated assessment model, demonstrating the link of discount rates and other aspects of global energy system modeling.

Beyond already considered aspects within investment risks, the impact of natural hazards driven by global warming are not estimated separately at all. Their sequence and magnitude are expected to increase in the future [29]. Countries that are currently safe for the installation of technology infrastructure and for people now, can be threatened by floods, hurricanes or other hazards in the future [30]. These arising risks are not yet considered explicitly in energy system models, although the discount rate is a tool for incorporating them.

While Damodaran's discount rate implicitly incorporates risks associated with past and present natural hazards beyond other political and economic impact factors by considering finance market values, it does not allow for their explicit consideration and analysis of their impact in increasing order and magnitude. Therefore, this paper aims to fill this gap by providing an approach to derive country-specific discount rates considering economic and natural hazard risk rates for each country in the world. This new approach is tested and applied for the case study of country-specific hydrogen production cost in the year 2050. With this newly developed approach, investment and annualized costs for energy system infrastructure and subsequently energy costs can be modeled in a more differentiated manner on a country-specific scale. It provides a robust base for specific investment decisions. Allowing for a wider application of the results, all discount rates derived from the approach are provided in the supplementary data of this paper.

# 2 Methodology

The methodology section is organized as follows. Chapter 2.1 presents the applied approach to calculating country-specific LCOH from given discount rates and cost parameters. The derivation of the applied discount rates including natural hazard risks is explained in section 2.2. Since the introduced equations in section 2.1 require numerous techno-economic parameters, the conduction of these parameters is further explained in section 2.3.

### 2.1. Overall approach to calculate hydrogen production costs

To determine the country-specific LCOH, an energy system model is established for each country globally. The general setup of the model is shown in Figure 1. Onshore wind power plants and horizontal single axis tracking PV power plants generate electricity to operate the electrolysis. The hydrogen produced is applied to meet a country-specific hydrogen demand. In general, the energy system models are set up with an hourly resolution throughout the year to accommodate variations in renewable potential. To overcome periods of low renewable electricity generation, hydrogen can be stored in gaseous hydrogen storage tanks. This allows the hydrogen demand to be decoupled from the renewable generation potential.

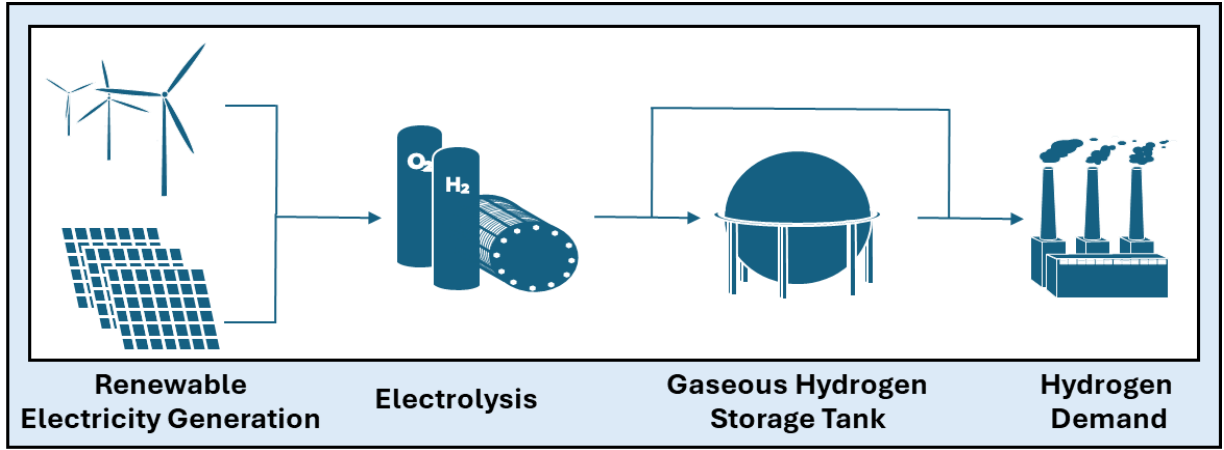

*Figure 1 Concept diagram for the comparison of hydrogen production costs*

Computationally, the energy system model framework ETHOS.FINE [31] is applied. The approach minimizes the total annual system costs to ensure a cost-minimal sizing of the technology capacities. Furthermore, a cost-minimal ratio of onshore wind power plants and PV plants is built to meet the hydrogen demand. According to Equation 1, the hydrogen production cost $LOHC_c$ for a country $c$ is calculated by dividing the sum of the technology-specific annualized investment cost $C_{inv,c,t}$ and the technology-specific annual operating cost $O_{c,t}$ for all involved technologies $T$ by the total annual hydrogen demand $D_{c,H_2}$.

$$LCOH_c = \frac{\sum_{t \in T}(C_{inv,c,t} + O_{c,t})}{D_{c,H_2}} \quad (1)$$

During this research, the country-specific discount rate $i_c$ for a given country $c$ is applied to decompose the overnight cost $I_{t,c,y}$ for a given technology $t$ and country into the annual investment cost $C_{t,c,y}$. In addition to the overnight cost $I_{t,c,y}$, the technology-specific economic lifetime $n_t$ is required as an input. Equation 2 leads to the required annuity cost $C_{t,c,y}$.

$$C_{t,c,y} = I_{t,c,y} * \frac{i_c(1+i_c)^{n_t}}{(1+i_c)^{n_t}-1} \quad (2)$$

The presented LCOH approach allows the comparison of different energy systems to identify favorable locations for hydrogen production. As well as the economic lifetime, the applied discount rate has a significant impact on the resulting annualized investment cost and, by extension, the LCOH.

## 2.2 Incorporation of economic and natural hazard risk to final discount rate

Among other factors, an increased investment risk results in an increased discount rate. As described in the introduction, global projections of LCOE and LCOH regularly apply either constant discount rates or relate to financial market data and consequently to investment risks applying the approach by Damodaran [26], [32], [33]. Another approach to determine the discount rate is the (WACC). In contrast to the Damodaran approach, the calculation of the WACC requires detailed financial information. To bypass these data, research often applies the above-mentioned approach of Damodaran, which processes available financial market data to an appropriate discount rate.

$$i_c = i_{e,c} + i_{n,c} \qquad (3)$$

Within this paper, the country-specific discount rate $i_c$ for a country $c$ is calculated as shown in Equation (3) by summing up the economic discount rate $i_{e,c}$ and a discount rate that includes the risk of natural hazards $i_{n,c}$. These discount rates are referred to as the economic discount rate and the natural hazard discount rate, whereas the resulting discount rate $i_c$ is referred to as the final discount rate. Natural hazards represent one of the most significant risks to energy system infrastructure, both in the present and in the future. Thus, the natural hazard risk is incorporated as a standalone factor in the final discount rate, as outlined in the proposed approach.

The country-specific final discount rate is determined for all countries globally. Following the convention of the ISO-3166 standard [34], the discount rates are determined for 254 countries. Due to its assumed negligible relevance, Antarctica as a defined country is excluded from this analysis.

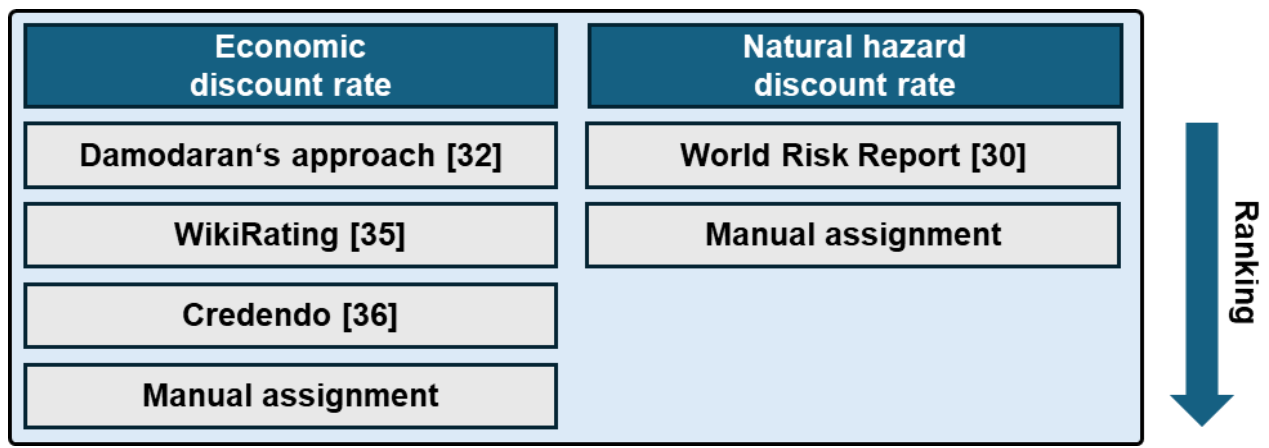

*Figure 2. Ranking of input datasets for derivation of economic and natural hazard discount rates*

Figure 2 indicates the applied datasets to determine both the economic and the natural hazard discount rate. Thus, the country-specific economic discount rate $i_{e,c}$ considers financial market data to translate the risks into a discount rate applying Damodaran's approach. In this research, the economic risks are adapted from Damodaran's published values [26], [32], [33]. Although Damodaran describes up to six different approaches to determine country risk rates, the course of this paper follows his recommended approach applying credit default swaps for 10-year government bonds and compares them to risk-free markets. In this way, Damodaran considers the country-specific political and economic stability. These economic values are available for multiple years and allow an analysis of country-specific risk rate evolution. Although Damodaran has continuously added more countries to his published results, his publications do not cover all countries according to the previously mentioned ISO 3166 code. For those countries, the WikiRating [35], is applied. This index also ranks countries with the help of rating agencies, and the approach is like Damodaran's approach as well. Thus, the rating is suitable to be transferred to Damodaran's rating approach. After applying these combined approaches, some countries are still not assigned a discount rate. These missing country-specific discount rates are calculated by applying the Credendo score [36]. The Credendo scale of 1 to 7 is therefore transformed into the same scale as for Damodaran applying a discrete scale of 21 values. Both the Credendo score and the WikiRating are only available for the most recent update, currently in 2024. For countries that are assigned by applying one of these approaches, the evolution of the risk over time cannot be analyzed. After that, there are still 14 small island countries, such as the Cook Islands, the Christmas Island, South Georgia and Sandwich Islands or political disputed islands, such as the Spratly Islands and the Clipperton Islands, that do not have an assigned country risk. These countries are manually assigned either to a comparable country's value or Damodaran's worst available rating.

Besides the exposure of countries to natural hazards itself, the vulnerability of societies and the ability of countries to cope with the consequences of those natural hazards is an important factor for determining the risk of natural hazards [37]. Thus, the World Risk Report [30] is considered as the most appropriate to determine the country-specific natural hazard risk rate $i_{n,c}$. This dataset covers 192 of the defined countries and ranks them in several categories regarding the exposure factors, covering the natural hazard risk and the vulnerability factor. The rating scale ranges from 0 to 100 and is published annually. To translate the World Risk Report scale to the Damodaran scale, the maximum rating value of the World Risk Report is normalized to the maximum rating value of the economic discount rate. Countries that are not analyzed in the World Risk Report are manually assigned to risk rates of neighboring countries. These missing countries are mostly small islands. The supplementary data includes an overview of the applied data assignment per country for both the economic and natural hazard discount rate

$$i_c = a * i_{e,c} + b * i_{n,c} \qquad (4)$$

$$a + b = 1 \quad \text{where } 0 \leq a \leq 1 \text{ and } 0 \leq b \leq 1 \qquad (5)$$

To avoid an unrealistically high discount rate and consequently high total annual energy infrastructure costs, the maximum final discount rate is limited to the maximum economic discount rate by applying equation 4 and equation 5. Thus, the final discount rate does not exceed Damodaran's discount rate that is already proven scientifically. Additionally, the ratio of the economic discount rate to the natural hazard discount rate can be adjusted to provide a variation in the final discount rate.

## 2.3 Investment cost parameter

The calculation of country-specific LCOH requires exogenous cost parameters. All cost parameters are given in USD and, if necessary, are adjusted to the base year 2023 by applying relevant inflation rates [38]. In this analysis, power generation technologies are limited to open-field PV and onshore wind power plants by trends for the year 2050 because these technologies are both based on renewable power and the most promising in investment cost. To consider the regional differentiation of overnight investment costs for power generation technologies, the regional cost differentiation published by the IEA in its Net Zero Emission scenario [39] is used. The IEA only publishes cost assumptions for several countries or regions. This region-specific cost information is assigned to the countries according to the ISO-3166 standard. For some regions, values for multiple countries are published. Thus, a cost average for those regions is created to process multiple cost data per region. The specific investment cost assumptions shown in Table 2.

| Region | Overnight investment cost [$USD_{2023}$/kW] | | Annual Operation cost [% of overnight investment] | |
|---|---|---|---|---|
| | Onshore wind power plant | PV power plant | Onshore wind power plant | PV power plant |
| Africa | 1,630 | 510 | 2.6 | 3.7 |
| Asia Pacific | 1,728 | 479 | 2.6 | 3.2 |
| Central and South America | 910 | 312 | 2.6 | 3.3 |
| Eurasia | 1,426 | 635 | 2.6 | 3.6 |
| Europe | 1,614 | 427 | 2.6 | 2.9 |
| Middle East | 1,666 | 229 | 2.6 | 3.6 |
| North America | 1,114 | 458 | 2.8 | 3.6 |
| China | 999 | 291 | 2.5 | 3.6 |
| India | 999 | 250 | 2.7 | 3.3 |
| Japan | 3,185 | 895 | 2.6 | 2.6 |

*Table 2: Regional specific overnight investment costs for electricity generation technologies in 2050 adapted from IEA [39]*

In addition to the regionalized investment costs for power generation technologies, other techno-economic parameters are processed in the calculation of country-specific LCOH. These parameters are summarized in Table 3, including techno-economic information on electrolysis, as well as lifetime and operating cost data for the power generation technologies. The overnight investment cost for China is 330 $USD_{2023}$/kW, whereas this cost category is 470 $USD_{2023}$/kW for all other countries [39]. Further data included in Table 3 is valid globally without any country-specific differentiation. Potential water supply costs are negligible [40].

| Cost category | Value |
| --- | --- |
| Electrolysis overnight investment cost in 2050 [$USD_{2023}$ / kW] | 330 - 470 |
| Electrolysis operation cost [% of investment cost] | 3 |
| Electrolysis efficiency [$kWh_{hydrogen}$ / $kWh_{electricity}$] | 0.7 |
| Electrolysis economic lifetime [a] | 10 |
| Hydrogen storage tank overnight investment cost in 2050 [$USD_{2023}$ / kWh] | 20 |
| Hydrogen storage tank operation cost [% of investment cost] | 2 |
| Hydrogen storage tank economic lifetime [a] | 30 |
| Hydrogen storage charge efficiency | 0.98 |
| Hydrogen storage discharge efficiency | 0.998 |
| Lower heating value hydrogen [kWh/kg] | 33.33 |
| Onshore wind power plant economic lifetime [a] | 20 |
| PV power plant economic lifetime [a] | 20 |

*Table 3: Applied country-independent techno-economic parameters [39], [41]*

The annual country-specific hydrogen demand is defined by 25% of the total hydrogen production potential of a country, as this approach has been shown to be suitable for the analysis of hydrogen production costs [42]. The potential for the applied electricity generation technologies is based on a comprehensive land eligibility analysis combined with a subsequent disaggregation of the wind power, solar radiation[43] and on analyses carried out in collaboration with the International Energy Agency for its Global Hydrogen Review 2024 [44].

## 3 Results

The result section is divided into three main parts to illustrate the impact of natural hazard risk to LCOH. The first section illustrates the advantages of using multi-year average Damodaran values for the LCOH calculation, rather than using the most recent Damodaran values to determine the discount rate, as is done in the literature and as outlined in the introduction. The second section shows the resulting LCOH for the inclusion of natural hazard risk with varying proportions to the final discount rate and illustrates the differences occurring by its application. In addition, as outlined above, global uniform discount rates also play a role in literature. The difference to the newly developed discount rate approach in terms of the resulting LCOH is shown in section 3.3.

## 3.1 Definition of the appropriate period to determine the economic discount rate

The discount rates that Damodaran derives from financial market parameters vary widely from year to year. *Figure 3* illustrates the annual distribution of all available discount rates. Before the financial crisis in 2009, the distribution of discount rates is relatively stable, although there are some outliers. However, the distribution of discount rates is more volatile after the financial crisis than before, emphasizing the need for considering longer-term developments for investment decisions in energy systems given their often-long-lasting economic lifetime. Although the interquartile range decreased in the following years after 2009, there is a more volatile mean of the discount rate and a tendency towards a more volatile mean and interquartile range until 2024. Interestingly, the Covid-19 pandemic is not visible in the overall statistics. However, the values from the year 2023 lead to the highest interquartile range of 0.098 within the analyzed period. Even in 2024, a high interquartile range of 0.083 is observed, accompanied by a high mean of 0.104 and a median of 0.082. The increased discount rates in the year 2023 and 2024 could be the consequence of the war in Ukraine and the resulting natural gas crisis. It is important to note that, according to the chosen Damodaran approach, the available discount rates for all countries are aggregated to 21 values per year. These values differ from year to year. Therefore, a displayed statistical outlier, shown in Figure 3, could include more than one country. Taking 2024 as an example, the outliers are Belarus, Lebanon, and Venezuela with a country-specific discount rate of 0.28.

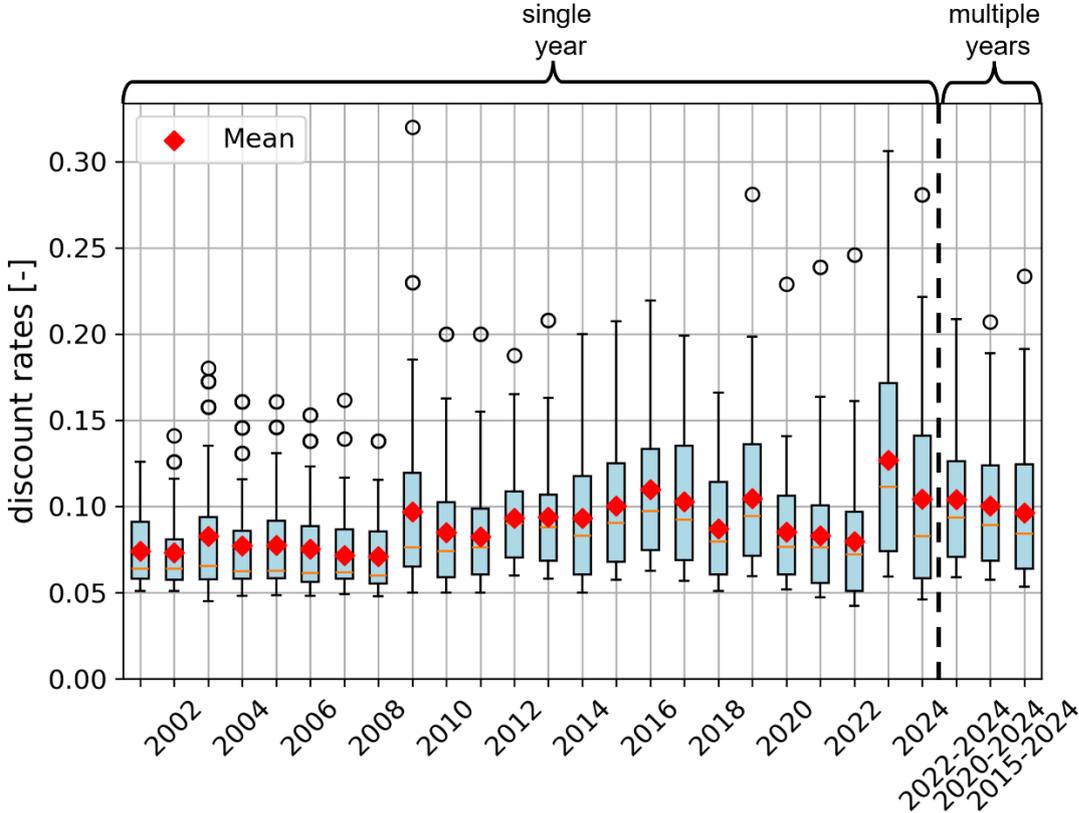

*Figure 3: Boxplots of Damodaran's annual discount rates from 2001 until 2024 [26] and the latest averaged discount rates for three, five and ten years*

Figure 4 illustrates a histogram of the country specific differences between the maximum and the minimum discount rate value within the observed period, proving a high volatility of Damodaran's discount rate for many countries. Therefore, it is inappropriate to focus on a single year as a reference for setting discount rates. To even out the effects of annual country-specific discount rates, deriving averages over certain time periods is required. However, the most appropriate timespan must be determined for the purpose of this paper. The results of the three-, five and ten-years-averages of the discount rates are depicted in Figure 3. In general, the more years are included in the derivation of the discount rate, the closer the mean is to the median. The mean value as well as the median value of the discount rate decreases, which is natural when comparing the discount rate values for the year 2024 with the corresponding time periods for the flattened values.

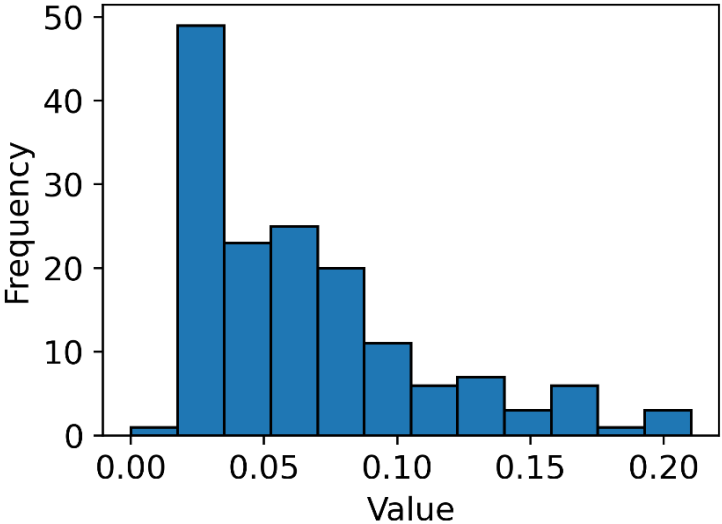

*Figure 4: Histogram of country-specific range between the maximum and minimum discount rate value per country between 2001 and 2024*

The annual discount rates from 2024 deviate by 0.78 percentage points from the ten-years average. Moreover, the standard deviation of the 2024 values is 1.67 percentage points higher than the corresponding value from the ten-years average. Given the comparatively low mean value and relatively high interquartile range among the averages discount rate approaches, a ten-year average is considered as the most suitable to smooth-out short-term events in the discount rate. It is therefore applied in the following sections. For those countries possessing no values for the last ten years, the appropriate discount rates are derived according to section 2.2. A comprehensive overview of the statistics for the analyzed timeframe is included in the supplementary data.

To illustrate the consequences that potentially occur by the application of either the annual discount rates from 2024 or the ten-year-average, the resulting LCOH are calculated. The calculations apply PV power generation, infrastructure cost such as PV generation technology and electrolysis assumptions for the year 2050. The calculation does not include countries that do not possess comprehensive data input within the last ten years.

The difference between the LCOH applying a ten-year average discount rate and the latest discount rates from 2024 is calculated as the delta LCOH and is shown in Figure 5. It is important to mention that the maximum and minimum values are set to 2 USD$_{2023}$/kg hydrogen and -2 USD$_{2023}$/kg hydrogen, respectively, to provide a suitable form of presentation. The calculated values may be outside the applied scale. A negative delta LCOH in blue colors indicates that the application of the annual discount rate approach for the year 2024 results in an increased LCOH compared to the flattened ten-year discount rate approach reaching from 2015 to 2024. This is directly connected to higher discount rates for the year 2024 than for the flattened ten-year-approach. For instance, countries in the southern part of South America and Southern Africa yield lower LCOH on the flattened ten-year-average discount rate approach. Belarus, Sri Lanka, and Ukraine are some of the countries with the highest delta LCOH of -5.64, -2.27 and -1.82 USD$_{2023}$/kg hydrogen. North America, Europe and Southeast Asia including Australia have positive deltas regarding their LCOH and therefore yield lower LCOH from the application of the annual discount rate from 2024 compared to the ten-years average discount rate approach. The most benefiting countries are Kyrgyzstan, Greece, and Jersey with a delta LCOH of 2.43, 1.73 and 0.83 USD$_{2023}$/kg hydrogen. The country-wise delta LCOH is shown in the Supplementary Data.

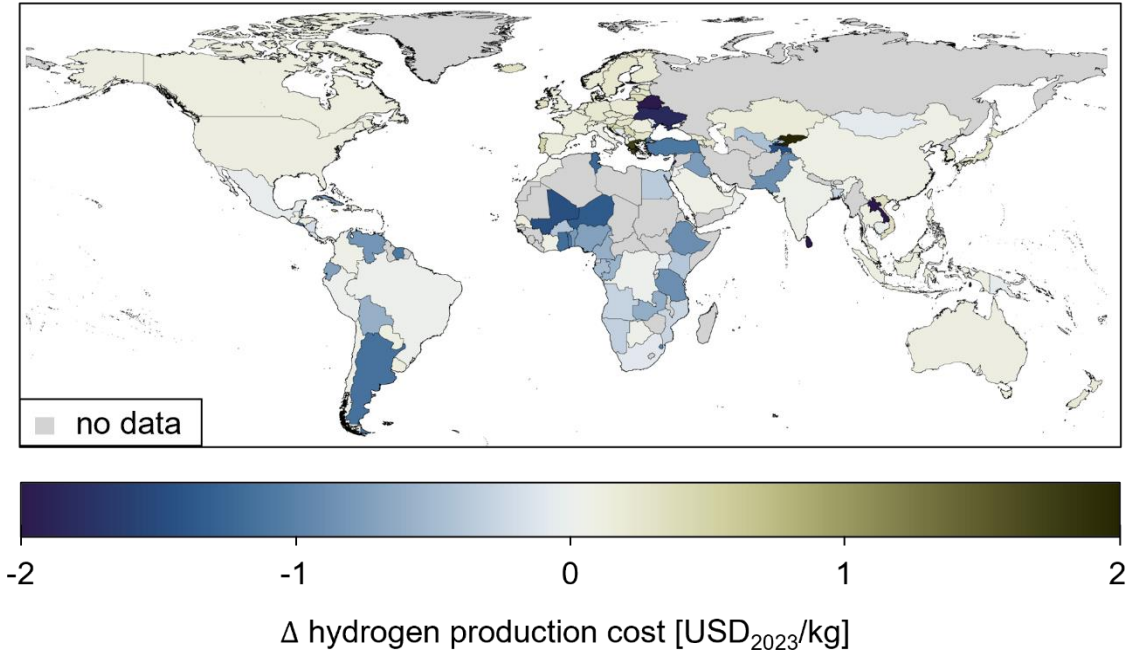

*Figure 5: Country-wise delta in hydrogen production cost in 2050 if the latest ten-year average of Damodaran's values is applied instead of its annual discount rate values from 2024*

The application of the economic discount rate by considering the evolution of the country-specific development within the last ten years leads to different LCOH compared to the application of a single year period as an input leading to a difference in LCOH up to 5.64USD$_{2023}$/kg hydrogen. It is important to mention that Damodaran's values are snapshots of a current economic and political situation, usually as of January 1$^{st}$ of each year. These values are heavily affected by the current political and economic situation of each country and therefore not perfectly suited for long-term consideration of LCOH.

Section 3.1 outlines the distinction between the application of annual discount rates and the application of average values over a specified period for all countries. To demonstrate the impact of climate change and risk on hydrogen production costs, this research employs an average discount rate over a ten-year period. The discount rates are derived from financial market data that is impacted by country-specific current political and financial developments. The selected period of ten years allows for compensation of short-term events that affect Damodaran's annual discount rate. Nevertheless, it should be noted that the whole analysis is based on historical data. Even an extension of the investigation period will not change the fact that the future development of finance market data remains uncertain. While the developed approach focuses on compensating for short-term events, there is a possibility of an overall change in the type of government and subsequent economic shifts in each country. For instance, Brazil has experienced such a change in the past. It is not feasible to anticipate these shifts by the application of historic data, particularly when considering future time horizons of twenty or thirty years. Accordingly, these changes are not covered by the methodology that is employed in this research.

### 3.2 The impact of natural hazard risks on hydrogen costs

As explained in Section 2.2, the final discount rate used to calculate the country-specific LCOH is a combination of the economic discount rate and the natural hazard discount rate. Depending on the proportion of these two risk rates in the final risk rate, the resulting LCOH will be different.

Figure 6 shows the derived economic and natural hazard discount rate for 254 countries in the world. In general, the Pearson correlation coefficient between all natural hazard discount rates and all economic discount rates is 0.31 with a p-value of 3.7e-7, which indicates a low positive correlation between the two discount rates. This is not surprising since Damodaran's economic discount rates must incorporate the natural hazard risks to a certain extent. Nevertheless, the low correlation between the two discount rates results in a final discount rate that differs from both input discount rates. Moreover, countries such as Argentina and Venezuela, which have a high economic discount rate of 16.9% and 23.3%, have a lower natural hazard discount rate of 7.7% and 11.1%, respectively. This fact illustrates the independence of the two factors. Therefore, it is appropriate to use them for the purpose of this research. The Figure 6 illustrates as well that European countries, Saudi Arabia, and Kazakhstan have relatively low discount rates for both the economic and the natural hazard risks and that countries with a comparatively lower economic discount rate, such as Mexico, Australia or India, do not necessarily have a low natural hazard discount rate. Both discount rates derived are listed in the supplementary data. To illustrate the impact of the discount rates on the modeled hydrogen production cost, the occurring LCOH by application of country-specific differentiated discount rates is analyzed.

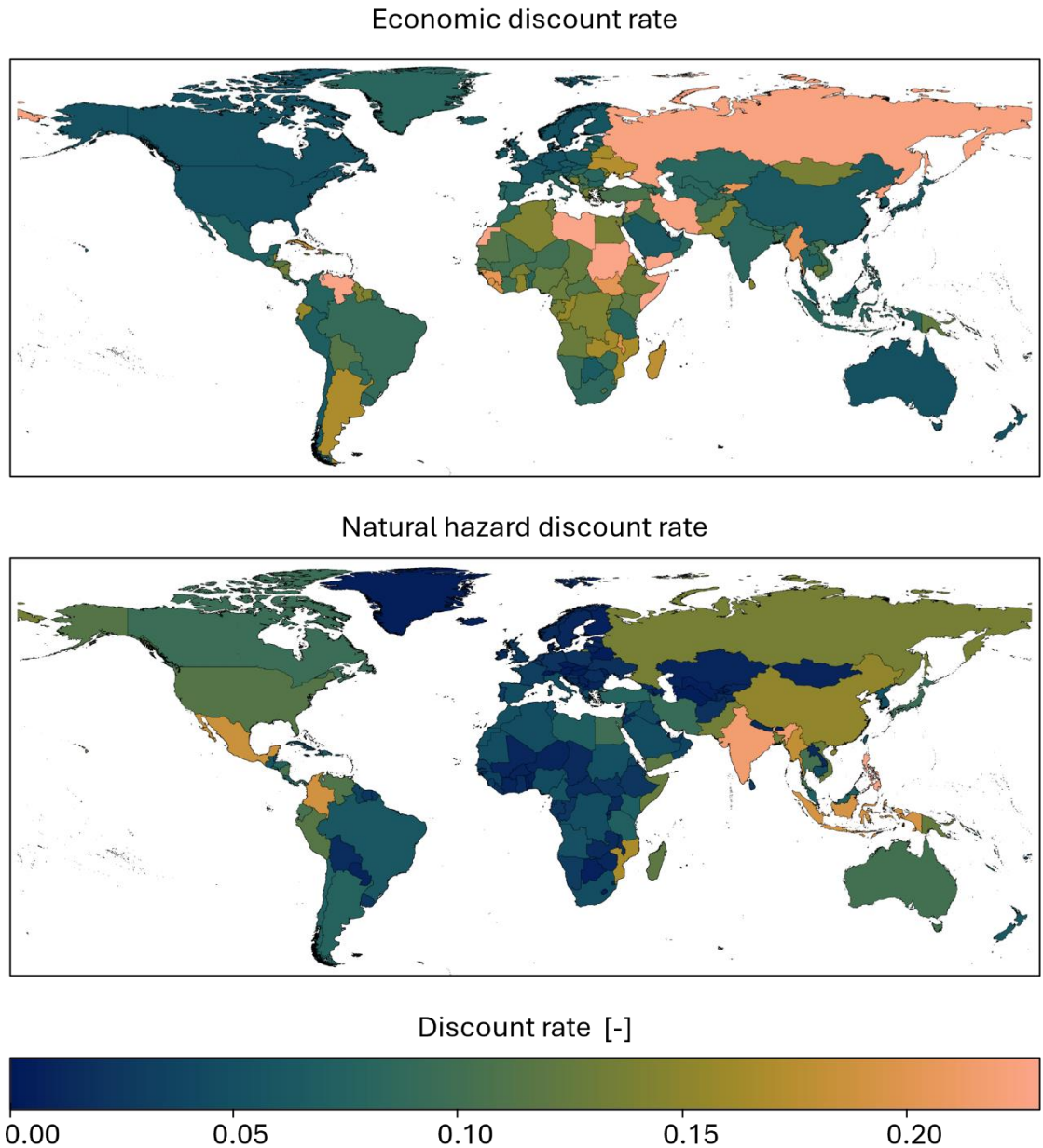

*Figure 6: Derived economic and hazard discount rate*

Figure 7 illustrates the country-specific LCOH in 2050 for an exclusive application of the economic discount rate and the natural hazard discount rate. The composition of the discount rates is based on a 10-year average, as outlined in Section 3.1. As can be seen in the upper part of Figure 7, a pure application of the economic discount rate under the given scenario conditions described above leads to the lowest LCOHs in Saudi Arabia (2.15 USD$_{2023}$/kg), United Arab Emirates (2.27 USD$_{2023}$/kg), Peru (2.34 USD$_{2023}$/kg) and Qatar (2.35 USD$_{2023}$/kg). However, countries such as Mayotte (24.87 USD$_{2023}$/kg), North Korea (11.94 USD$_{2023}$/kg), Belarus (10.56 USD$_{2023}$/kg) and Kyrgyzstan (10.24 USD$_{2023}$/kg) have comparatively high LCOH. As the natural hazard discount rate is different from the economic discount rate, the country-wise LCOH by the pure application of the natural hazard discount rate differs from the former presented LCOH.

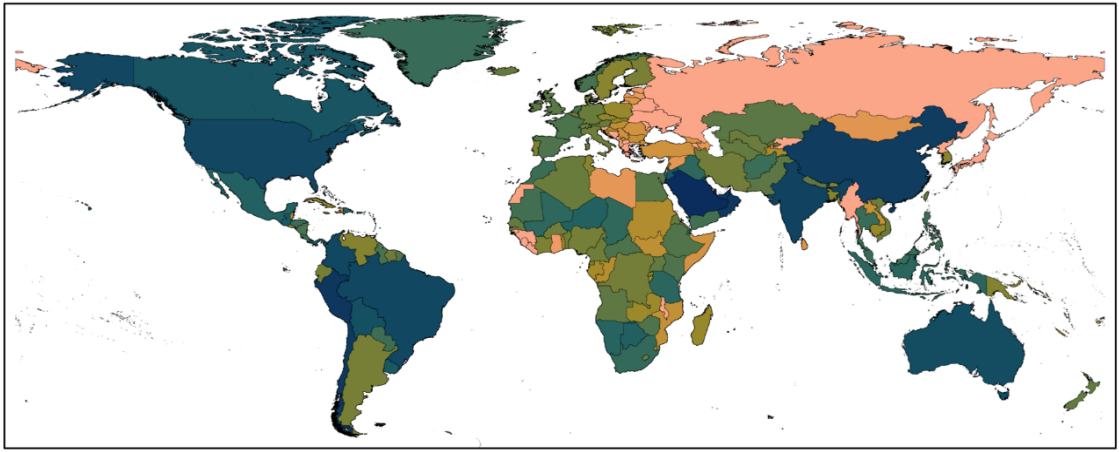

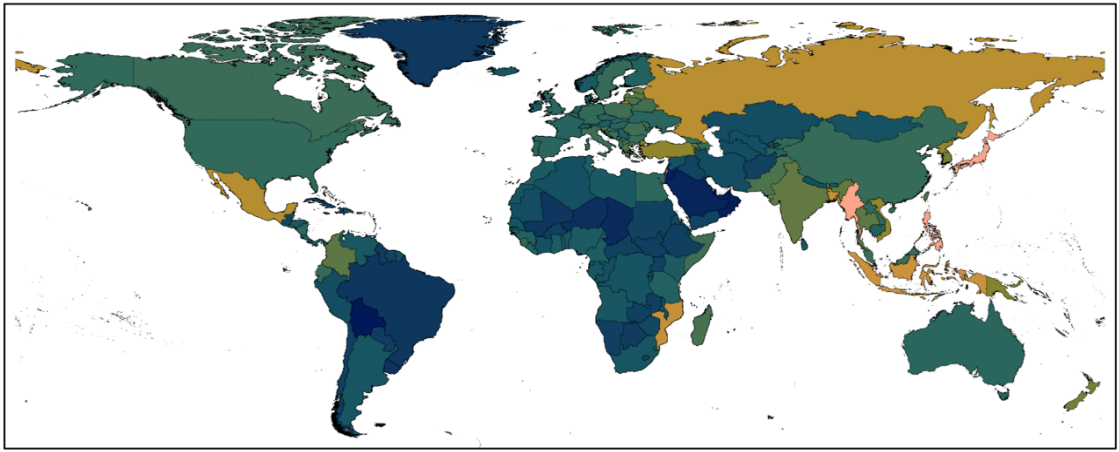

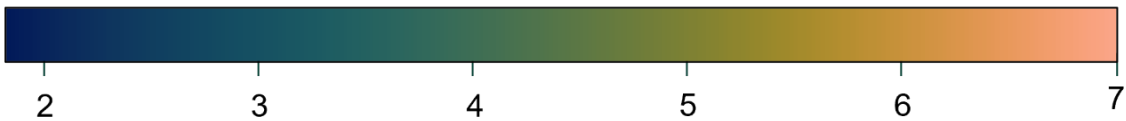

*Figure 7: Country-wise LCOH in 2050 for an exclusive application of economic discount rate and an exclusive application of natural hazard discount rate*

Since the other parameters relevant to the calculation of the LCOH remain constant, the effect of the change in LCOH is exclusively due to the change in the final discount rate composition. Figure 7 shows the LCOH that is derived solely by the application of the natural hazard discount rate as well. Following that, the lowest LCOH is now available in Qatar (1.77 USD$_{2023}$/kg), Bolivia (1.81 USD$_{2023}$/kg), Aruba (1.85 USD$_{2023}$/kg) and Bahrain (1.87 USD$_{2023}$/kg), whereas the LCOH is still comparatively high for other countries, such as Mayotte (13.39 USD$_{2023}$/kg), Japan (8.46 USD$_{2023}$/kg), Hong Kong (7.94 USD$_{2023}$/kg) and the Philippines (7.83 USD$_{2023}$/kg). Interestingly, the natural hazard risk dominated final discount rate benefits the African countries, leading to lower LCOH compared to a pure application of economic discount rates. Considering the fact that African countries are set among the countries to suffer the most from global warming due to extensive drought periods [45], [46], the risk of natural hazards is still below the impact of the economic risks and therefore a favorable proportion of the final discount rates. In contrast, Northern America is disadvantaged by the inclusion of natural hazard risks leading to an increased LCOH, while Qatar remains a low-cost producer of hydrogen in the analyzed parameter variations, regardless of the applied final discount rate.

For a better comparison, the relative cost difference of the application of both discount rates is shown in Figure 8. A positive relative difference in green indicates a higher LCOH if the natural hazard discount rate is applied instead of a pure economic discount rate. Countries with the highest positive relative difference are the Philippines (96.7% or 3.85 USD$_{2023}$/kg), India (74.3% or 1.95 USD$_{2023}$/kg), Mexico (66.4% or 2.28 USD$_{2023}$/kg), and Indonesia (65.9% or 2.39 USD$_{2023}$/kg). In contrast, the countries with the lowest negative values are Kyrgyzstan (-62.6% or -6.41 USD$_{2023}$/kg), Belarus (-61.5% or -6.50 USD$_{2023}$/kg), Malawi (-59.3% or -4.16 USD$_{2023}$/kg) and Guinea-Bissau (-58.2 or -4.84 USD$_{2023}$/kg). The second d value in brackets indicates the country-specific absolute difference in LCOH.

Countries indicating a relative difference close to zero prove a strong resilience against the impact of natural hazard risk to their LCOH and cost for infrastructure in general. Among the countries that were mentioned as low-cost hydrogen producers, Saudi-Arabia indicates with a relative difference of -7.2% the highest resilience against the inclusion of natural hazard risk.

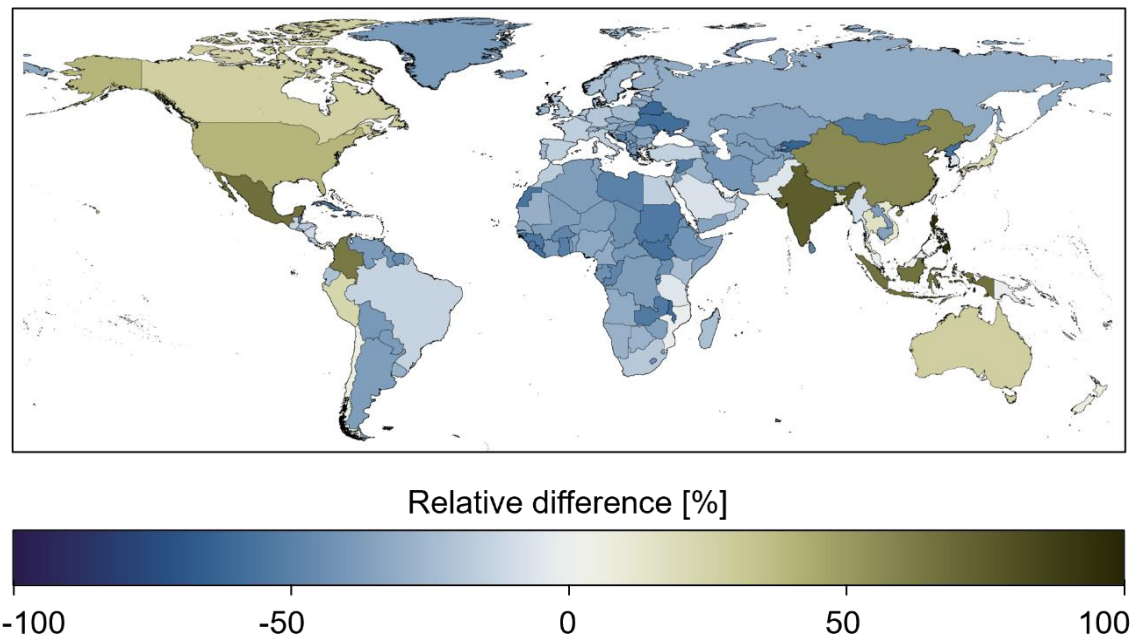

*Figure 8: Country-wise relative difference of LCOH in 2050 if applying the pure natural hazard discount rate instead of the pure economic discount rate*

The country-specific difference in LCOH creates a solution space that arises when selecting a specific ratio of the economic to the natural hazard discount rate. In general, North America, Southwest Asia, the western part of South America, China and Australia show lower LCOH when a higher proportion of the economic discount rate is applied to the final discount rate, whereas Europe, Northern Asia, Greenland, and Africa take advantage of applying a higher proportion of the natural hazard discount rate compared to the economic one in the final discount rate. The Asia-Pacific region is prone to natural hazards, including underwater earthquakes and storms, which have an impact on the region's natural hazard discount rate. Furthermore, tectonic plates building the San Andreas Fault and the frequently occurring hurricanes threaten Northern and Central America. As a result, the final discount rates are significantly influenced by the natural hazard discount rate. Notably, African countries are affected by economic rather than natural hazard discount rates. Given that the economic discount rate is already high and subject to political and financial instabilities, the addition of natural hazard risk does not have a detrimental impact on the final discount rates.

Considering equation 2 and equation 3, the final discount rate is the only parameter that countries can influence in terms of the calculation of LCOH. The economic discount rate of a country is primarily determined by financial market parameters, particularly by the risk of default on government bonds. A stable political environment over an extended period and a robust, globally integrated economy has a favorable impact on a country's risk rating, consequently reducing the risk of default. Thus, stable political and economic environments are associated with lower economic discount rates. The natural hazard discount rate consists of two primary inputs: exposure to natural hazards and the vulnerability of a country to such risks. The geographical location of a country is an inherent risk factor that cannot be changed. However, measures can be taken to decrease a country's vulnerability to certain natural hazards. For instance, the construction of dyke can decrease the risk of flooding in a defined area. Nevertheless, there are natural hazards whose impact on a country is difficult to mitigate. To decrease a country's potential vulnerability to earthquakes, the way of building construction must be adapted. This is feasible for planned energy system infrastructure, but more challenging for already existing ones and often connected to higher costs.

Regarding equations 4 and 5, different proportions of economic and natural hazard can be applied to derive the final discount rate. After considering the range of potential solutions for country-specific LCOH and the minimal correlation between economic and natural hazard discount rates, a ratio of 75:25 for economic to natural hazard discount rate is identified as the most suitable for calculating the final discount rate in this paper.

The final discount rates, which are conducted using different proportions, and the described absolute and relative difference in hydrogen generation cost in the event of applying either a pure economic or a pure natural hazard discount rate are included in the supplementary data.

### 3.3 LCOH gap for uniform discount rates
As outlined in section 3.1 and section 3.2, the application of an economic and natural hazard discount rate derived through analysis of values over a ten-year period affects the LCOH. Although, global uniform discount rates have been applied in the literature for calculating the country-specific LCOH. Figure 9 demonstrates the discrepancy in country-wise LCOH in 2050 between the application of a uniform discount rate and the developed final discount rate approach, which incorporates a final discount rate with a share of 75% economic and 25% natural hazard discount rate. As previously stated, other scientific work often employs an 8% uniform discount rate, which is set as quantity in this chapter.

The way the results are presented in Figure 9 is aligned with the concept depicted in Figure 8. A presented relative difference indicates the difference in LCOH resulting from the application of the derived discount rate instead of the global uniform discount rate. As a result, a highly differentiated impact of the application of the new final discount rate approach occurs globally. In general, a country-specific final discount below the uniform discount rate of 8% leads to a blue colored negative relative LCOH difference. Thus, highly developed countries in Western-Europe that possess neither high natural hazard risks nor a high economical risk potential can produce hydrogen to lower cost and are disadvantaged by the application of a global uniform discount rate of 8%, whereas a major part of Africa benefits from the application of the uniform discount rate.

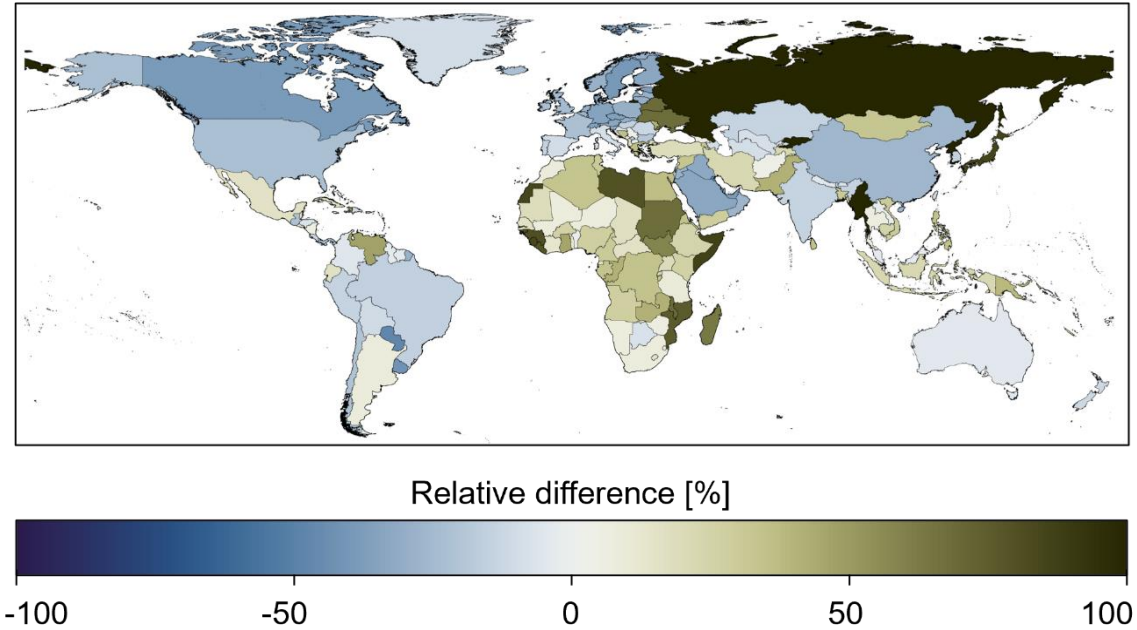

*Figure 9: Country-wise relative difference of LCOH in 2050 if applying the final discount ratio consisting of a 75% economic and 25% natural hazard discount rate instead of a uniform discount rate of 8%*

The discrepancy in LCOH is significant, contingent on the range between the uniform discount rate and the country-specific final discount rate. The application of the described approach in contrast to the uniform discount rate results in an increase in LCOH by 86.3% in Somalia, 83.7% in Russia, 79.3% in Myanmar, and 76.5% in Western Sahara. This equates to an absolute increase in LCOH of 2.57, 3.49, 3.52, and 4.42 USD $_{2023}$/kg hydrogen, respectively. However, several countries stand to gain significantly from the application of the conducted discount rate approach compared to the uniform discount rate application. These include Qatar with a reduction of 36.7%, Kuwait with 35.7%, the United Arab Emirates with 34.1%, and Saudi Arabia with 32.3%, or 1.27, 1.53. 1.12 and 1.01 USD$_{2023}$/kg. Please refer to the Supplementary Data for the country-wise calculated absolute and relative differences.

The results of this comparison demonstrate that failing to consider the influence of a country-specific discount has varying implications for the interests of different countries. For countries in Africa, Southeast Asia, Russia, and Mexico, among others, applying a global uniform discount rate of 8% results in an underestimation of the LCOH compared to the country-specific discount rates in the described scenario. Conversely, in Europe, the Middle East, China, Australia, Brazil, India, and other countries, the LCOH increased. The incorporation of economic and natural hazard risk may impact the interests of certain countries.

# 4 Conclusion

Future energy infrastructure will be exposed to increasing risks from natural hazards. These risks are distributed unevenly globally and need to be incorporated into the decision-making process for energy infrastructure investments. Therefore, this paper provides a new approach to derive country-specific discount rates focusing on natural hazard risks at a country-level. To exemplify the impact of the natural hazard risk inclusion, the LCOH is proven to be highly volatile regarding the proportion of natural hazard risk inclusion.

Despite the latest tendency of other researchers to apply the latest finance market data most often through Damodaran's approach to translate these finance market data into country-specific discount rates, our research demonstrates that considering a ten-year-period to derive the mean of those values is a more suitable approach in a master planners view for building energy system infrastructure. Moreover, the changing climate conditions and therefore, the increasing risks associated with natural hazards must be included in the discount rate to cover the future extended infrastructure investment risks. The proposed approach translates these risks into a combined discount rate and provides a solid base to incorporate them into the risk assessment for investment decisions. In general, the degree to which the natural hazard risk is considered within the investment calculation is variable and can be adapted by the user's need. However, the examples introduced demonstrate the significant influence that selected discount rates can have on country-specific LCOHs. A shift from pure economic discount rates to pure natural hazard discount rate could result in a country-specific decrease from -62.6% or -6.41 USD$_{2023}$/kg hydrogen in Kyrgyzstan to an increase of up to 96.1% or +3.85 USD$_{2023}$/kg hydrogen in the Philippines of the LCOH in the year 2050. Furthermore, countries, such as Suai-Arabia or Mozambique. These results are already noteworthy when considered as a standalone value. Some countries, such as Saudi-Arabia, are not affected by the introduction of natural hazard risks in the final discount rate.

Theoretically, measures are available to impact the discount rate. Potential default risks of country-specific government bonds primarily affect the economic discount rate as one part of the final discount rate approach. A stable political and economic environment of a country will improve the economic discount rate and, thus, the final discount rate. Therefore, appropriate policy measures are required to improve the economic discount rate. The country-specific natural hazard risk is divided into two parameters. The first is determined by the geographical location of a country and cannot be changed by its nature. The second is a country's vulnerability of a country to a nature hazard. Appropriate measures can be taken to improve this parameter and thus improve the natural hazard discount rate. However, it is important to note that for certain natural hazards, vulnerability reduction may be difficult.

The approach to incorporating natural risks demonstrates the substantial impact of future climate change risks on investment decisions, highlighting the necessity for consideration of both environmental and economic factors in investment decision-making. As LCOH is frequently utilized as a decision criterion in energy system transformation pathways, the inclusion of future natural hazard risks can influence the outcomes and, consequently, the policy recommendations. While this research focuses on the impact of including country-specific LCOH, the application of the discount rate approach is not limited to this. This approach can be applied in several other areas where future natural hazard risk will have an impact on investment decisions, such as further energy system infrastructure, or the construction of private and company buildings. The required country-specific discount rates for both the economic and the natural hazard discount rate are published alongside the different proportions of them in a final discount rate in the supplementary data.

# Declaration of competing interest

The authors declare that they have no known competing financial interest or personal relationships that could have appeared to influence the work reported in this manuscript.

# Acknowledgements

This work was supported by the Helmholtz Association under the program "Energy System Design" and funded by the European Union (ERC, MATERIALIZE, 101076649). Views and opinions expressed are, however, those of the authors only and do not necessarily reflect those of the European Union or the European Research Council Executive Agency. Neither the European Union nor the granting authority can be held responsible for them.

# Author Contribution